\documentclass[10pt]{revtex4}
\usepackage{hyperref}
\usepackage{color}

\begin{document}
\title { ABJM theory
in Batalin-Vilkovisky formulation}
  
 \author { Sudhaker Upadhyay  }
 \email{ sudhakerupadhyay@gmail.com}
\address{Departamento de F\'{\i }sica Te\'{o}rica, Instituto de F\'{\i }sica, UERJ - Universidade do Estado do Rio de Janeiro,
 \\ \small \textnormal{ \it Rua S\~{a}o Francisco Xavier 524, 20550-013 Maracan\~{a}, Rio de Janeiro, Brasil}\normalsize.}
 
 \author{Diptarka Das}
 \email{diptarka.das@uky.edu}
 \address{Department of Physics and Astronomy, \\ 
 University of Kentucky, Lexington, KY 40506, USA.}
\begin{abstract}
 We analyze the quantum ABJM theory on ${\cal N}=1$ superspace in different gauges.
 We study the Batalin-Vilkovisky (BV) formulation for this model. By developing field/antifield
 dependent BRST transformation we establish connection between the two
 different solutions of the quantum master equation  within the BV formulation.  
\end{abstract}

\maketitle
\section{Introduction}
 The Aharony-Bergman-Jafferis-Maldacena (ABJM) theory is a conformal field theory in three dimensional spacetime.
The ABJM  theory with gauge group  $U(N)\times U(N)$ 
is  represented by $N$ $M2$-branes and has been constructed recently \cite{abjm, 1a}. 
More precisely, it is shown that $\mathcal{N} = 6$ supersymmetric Chern-Simons quiver
gauge theory with bifundamental matter enjoying $SO(4)$ flavor symmetry is dual to $M$-theory compactified on $AdS_4 \times S^7/Z_k$, and describes the low energy dynamics of a stack
of $M2$-branes probing an orbifold singularity.  This theory only has 
  $\mathcal{N} = 6$ supersymmetry but it is expected
to be enhanced to the full $\mathcal{N} = 8$ supersymmetry \cite{abjm2}. The M2-brane branes ending on M9-branes and gravitational waves have also been studied
  \cite{ber}.
  
   It may be noted that as the ABJM theory has gauge symmetry, it cannot be quantized without getting rid 
of these unphysical degrees of freedom. This 
can be done by fixing a gauge. The gauge fixing condition can be incorporated at a quantum level by 
adding  ghost and gauge fixing terms to the original classical Lagrangian. It is known that 
for  a gauge theory the new effective Lagrangian 
constructed as the sum of the original classical Lagrangian with the gauge fixing and the ghost terms, 
is invariant under a new set of transformations called the BRST transformations \cite{brst, brst1}.
 BRST symmetry has also been studied in non-linear gauges  \cite{nlbrst,nlbrst1}.
 
 On the other hand the field/ antifield formulation also known as the Batalin-Vilkovisky (BV) formalism \cite{henna} -\cite{bv3} is one of the most powerful techniques to study gauge field theories. 
 The generalization of BRST by making the infinitesimal BRST parameter finite
 and field-dependent, known as FFBRST formulation  \cite{sdj} has many application in gauge field theories \cite{sdj,sdj1,rb,susk,jog, sb2,smm,fs,sud1,rbs,sudhak,rs}.
 Recently, we generalize the BRST symmetry by making the parameter
 field/antifield dependent for super-Chern-Simons theory \cite{sudhak0}.
 We generalize  such formulation in the case of ABJM theory on ${\cal N}=1$ superspace in the BV formalism.
  
In this work we discuss the ABJM theory from the perspective of gauge theory
by discussing different gauge conditions. We investigate the different effective actions
corresponding to the different gauge choices. We 
establish the BRST symmetry for the theory using $two$ Grassmann parameters.
Furthermore, the general BV quantization  of the model has been analyzed. 
  We generalize the BRST symmetry of the model by making the parameters field/antifield 
  dependent. We compute the resulting Jacobian coming from the functional measure of the general generating functional. We find that for a particular $choice$ of field/antifield dependent parameters, (equations \ref{res1} and \ref{res2})
 the different gauges of ABJM theory can be connected.
  This result will be helpful to interrelate computations of physical quantities of the ABJM theory in linear and non-linear gauges.
  
  The paper is presented in following way.
  In Sec. II, we analyze the classical ABJM theory in ${\cal N}=1$ superspace from the gauge symmetric point of view. Sec. III is devoted to describe the 
  quantum analysis by studying different gauge conditions. 
  The BV formalism is developed for ABJM theory in section IV, which widens the quantization scheme.
  In Sec. V, we developed a mapping between different solutions of extended quantum action
  using the techniques of field/antifield dependent BRST symmetry.
  The results are summarized  in the last section. 
\section{The   ABJM theory in ${\cal N}=1$ superspace}

We start with the  Chern-Simons Lagrangian densities $\mathcal{L}_{CS}$, $\tilde{\mathcal{L}}_{CS}$ with gauge group's $U(N)_k$ and $U(N)_{-k}$ on ${\cal N}=1$ superspace defined by
\begin{eqnarray}
 \mathcal{L}_{CS}& =& \frac{k}{2\pi} \int d^2 \,  \theta \, \, 
 \mbox{Tr} \left[  \Gamma^a     \omega_a + \frac{i}{3} [\Gamma^a, \Gamma^b]_{ }
  D_b \Gamma_a  
 + \frac{1}{3} [\Gamma^a,\Gamma^b]_{ } [\Gamma_a, \Gamma_b]_{ }
\right] _|, 
\nonumber \\
 \tilde{\mathcal{L}}_{CS} &=& \frac{k}{2\pi} \int d^2 \,  \theta \, \, 
 \mbox{Tr} \left[  \tilde{\Gamma}^a     \tilde{\omega}_a
 + \frac{i}{3} [\tilde{\Gamma}^a, \tilde{\Gamma}^b]_{ }
  D_b \tilde{\Gamma}_a
+ \frac{1}{3} [\tilde{\Gamma}^a,\tilde{\Gamma}^b]_{ } 
[\tilde{\Gamma}_a, \tilde{\Gamma}_b]_{ }\right] _|, 
\end{eqnarray}
where $k$ is an integer playing the role of a coupling constant. $\omega_a$ and $\tilde \omega_a$ have following expression:
\begin{eqnarray}
 \omega_a &=&  \frac{1}{2} D^b D_a \Gamma_b - i  [\Gamma^b, D_b \Gamma_a]_{  } 
- \frac{2}{3} [ \Gamma^b ,
[ \Gamma_b, \Gamma_a]_{ }]_{ },\nonumber \\
   \tilde\omega_a &=&  \frac{1}{2} D^b D_a \tilde\Gamma_b -i  [\tilde\Gamma^b, D_b \tilde\Gamma_a]_{  } - 
\frac{2}{3} [ \tilde\Gamma^b,
[ \tilde\Gamma_b,  \tilde\Gamma_a]_{  } ]_{  }. 
\end{eqnarray}
The $D_a$ represents the super-derivative  defined as
\begin{equation}
 D_a = \partial_a + (\gamma^\mu \partial_\mu)^b_a \theta_b,
\end{equation}
and  $'|'$  means that the quantity is evaluated at $\theta_a =0$. 
In component form the gauge connections  $\Gamma_a$ and $\tilde \Gamma_a$ are expressed as 
\begin{eqnarray}
 \Gamma_a = \chi_a + B \theta_a + \frac{1}{2}(\gamma^\mu)_a A_\mu + i\theta^2 \left[\lambda_a -
 \frac{1}{2}(\gamma^\mu \partial_\mu \chi)_a\right], \nonumber \\
 \tilde\Gamma_a = \tilde\chi_a + \tilde B \theta_a + \frac{1}{2}(\gamma^\mu)_a \tilde A_\mu + i\theta^2 \left[\tilde \lambda_a -
 \frac{1}{2}(\gamma^\mu \partial_\mu \tilde\chi)_a\right]. 
\end{eqnarray}
The explicit expression for the Lagrangian density of the matter fields  is given by 
\begin{eqnarray}
 \mathcal{L}_{M} =\frac{1}{4} \int d^2 \,  \theta \, \,  
 \mbox{Tr} \left[ [\nabla^a_{(X)}           X^{I \dagger}           
\nabla_{a (X)}           X_I ] +
[\nabla^a_{(Y)}           Y^{I \dagger}           \nabla_{a (Y)}            Y_I ] + 
 \frac{16\pi}{k} \mathcal{V}_{    } \right]_|,
\end{eqnarray}
where 
\begin{eqnarray}
 \nabla_{(X)a}          X^{I } &=& 
D_a  X^{I } + i \Gamma_a          X^I 
-     i   X^I    \tilde\Gamma_a  , \nonumber \\ 
 \nabla_{(X)a}          X^{I \dagger} 
&=& D_a  X^{I  \dagger}      
+ i \tilde\Gamma_a        X^{I  \dagger}- 
     i X^{I  \dagger}   \Gamma_a, \nonumber \\ 
 \nabla_{(Y)a}          Y^{I } &=& D_a  Y^{I }  
+ i \tilde\Gamma_a          Y^I- i  Y^I   \Gamma_a , \nonumber \\ 
 \nabla_{(Y)a}          Y^{I \dagger} &=& D_a  Y^{I  \dagger} 
+ i \Gamma_a     
       Y^{I  \dagger} 
 - i   Y^{I  \dagger}     \tilde\Gamma_a.
\end{eqnarray}
Now, the  classical Lagrangian density for ABJM theory with the gauge group $U(N)  \times U(N) $ 
  on ${\cal N}=1$ superspace  is given by, 
\begin{equation}
{ \mathcal{L}_c} =  \mathcal{L}_{M} + \mathcal{L}_{CS} - \tilde{\mathcal{L}}_{CS},
\end{equation} 
which remains covariant under the following gauge transformations:
\begin{eqnarray}
\delta  \,\Gamma_{a} = \nabla_a   \xi, && \delta \, \tilde\Gamma_{a} =\tilde\nabla_a    
 \tilde \xi, \nonumber \\
\delta  \, X^{I } = i \xi X^{I }    -  iX^{I }  \tilde \xi, 
 &&  \delta  \, X^{I \dagger }
 = i   \tilde \xi  X^{I \dagger }  - i  X^{I \dagger } \xi, 
\nonumber \\
\delta  \, Y^{I } = i  \tilde \xi  Y^{I }  -iY^{I }  \xi,  &&  
\delta  \, Y^{I \dagger } = i \xi  Y^{I \dagger } -  i Y^{I \dagger }
  \tilde\xi,
\end{eqnarray}
with the local parameters $\xi$ and $\tilde\xi$. The super-covariant derivatives  $\nabla_a$ and $\tilde\nabla_a$ are defined by
\begin{equation}
\nabla_a =D_a-i\Gamma_a, \ \ \ \tilde\nabla_a =D_a-i\tilde\Gamma_a.
\end{equation}
\section{Gauge conditions and BRST   symmetry}
In this section, we investigate the quantum action for ABJM theory in linear and non-linear gauges.
The nilpotency of BRST symmetry is also demonstrated for this theory.
\subsection{Linear gauge}
Being gauge invariant,  the non-Abelian Chern-Simons theory on ${\cal N}=1$ superspace   contains some redundant degrees of freedom.
To quantize the theory correctly we need to choose a gauge.
The covariant (Lorentz-type) gauge fixing conditions for ABJM theory are
\begin{eqnarray}
G_1 \equiv D^a \Gamma_a  =0,\ \  \tilde G_1 \equiv D^a \tilde{\Gamma}_a =0.
\end{eqnarray}
These  gauge fixing conditions can be incorporated in the theory at the quantum level by adding the following  gauge fixing term to 
the original Lagrangian density,
\begin{equation}
\mathcal{L}_{gf} = \int d^2 \,  \theta \, \,  \mbox{Tr}  \left[ib_1   (D^a \Gamma_a) + \frac{\alpha}{2}b_1   b_1 -
i \tilde{b}_1    (D^a \tilde{\Gamma}_a) - \frac{\alpha}{2}\tilde{b}_1    \tilde{ b}_1
\right]_|,
\end{equation}
where $b_1$ and $\tilde b_1$ are the Nakanishi-Lautrup auxiliary fields. The Faddeev-Popov ghost terms corresponding to the above gauge fixing term is constructed as  
\begin{equation}
\mathcal{L}_{gh} = \int d^2 \,  \theta \, \,   \mbox{Tr}
\left[ i\bar{c}_1   D^a \nabla_a    c_1 -i \tilde{\bar{c}}_1    D^a \tilde{\nabla}_a   \tilde{c}_1 \right]_|.
\end{equation}
Now, we define the full quantum action for ABJM theory
in Lorentz-type gauge by writing the gauge-fixing and the ghost terms collectively with classical action
\begin{equation}
 {\cal L}_{L}={\cal L}_c +{\cal L}_{gf} + {\cal L}_{gh}. 
  \end{equation}
The BRST transformations, which leaves the above effective action  invariant, are written by
\begin{eqnarray}
\delta_b \,\Gamma_{a} = \nabla_a    c_1\ \Lambda, && \delta_b\, \tilde\Gamma_{a} =\tilde\nabla_a    
 \tilde c_1\  \tilde \Lambda, \nonumber \\
\delta_b\,c_1 = - {[c_1,c_1]}_ { } \Lambda, && \delta_b \,\tilde{ {c}}_1 = -   [\tilde{ {c}}_1 ,  \tilde c_1]_{ } \tilde \Lambda, \nonumber \\
\delta_b \,\bar{c}_1 = b_1\ \Lambda, && \delta_b \,\tilde {\bar c}_1 = \tilde b_1\  \tilde \Lambda, \nonumber \\ 
\delta_b \,b_1 =0, &&\delta_b \, \tilde b_1= 0, \nonumber \\ 
\delta_b \, X^{I } = i c_1   X^{I }  \Lambda -  iX^{I }  \tilde c_1 \tilde \Lambda, 
 &&  \delta_b \, X^{I \dagger }
 = i   \tilde c_1    X^{I \dagger }  \tilde \Lambda- i  X^{I \dagger }  c_1\ \Lambda, 
\nonumber \\
\delta_b \, Y^{I } = i  \tilde c_1   Y^{I } \tilde \Lambda -iY^{I }   c_1\ \Lambda,  &&  
\delta_b \, Y^{I \dagger } = i c_1   Y^{I \dagger }\ \Lambda -  i Y^{I \dagger }
  \tilde c_1\ \tilde \Lambda,
  \label{brstl}
\end{eqnarray}
where $\Lambda$ and $ \tilde \Lambda$ are the infinitesimal
anticommuting parameters of transformation. 
 \subsection{Non-linear gauge }
 We start this subsection by demonstrating the ABJM theory in non-linear gauge  as follows
 \begin{eqnarray} 
 {\cal L}_{NL}   &=& {\cal L}_c+ \int d^2\theta \ \mbox{Tr} \bigg[ \frac{\alpha}{2}  b_2^2  + i b_2  D^a \Gamma_a  -iD^a \bar{c}_2\nabla_ac_2   - \frac{i}{2} D^a \Gamma_a [\bar{c}_2,c_2]  \nonumber \\  &+&    \frac{\alpha}{8} [\bar{c}_2,c_2]^2 - \frac{\alpha}{2}b_2 [\bar{c}_2,c_2]   +  iD^a \tilde{\bar{c}}_2\nabla_a\tilde{c}_2 - \frac{\alpha}{2} \tilde{b}_2^2   - i\tilde{b}_2  D^a \tilde{\Gamma}_a  \nonumber \\ 
 &+&  \frac{i}{2} D^a \tilde{\Gamma}_a [\tilde{\bar{c}}_2,\tilde{c}_2]- \frac{\alpha}{8} [\tilde{\bar{c}}_2,\tilde{c}_2]^2 +\frac{\alpha}{2} \tilde{b}_2 [\tilde{\bar{c}}_2, \tilde{c}_2] \bigg]_|.
\end{eqnarray}
The above Lagrangian density can be obtained by performing the following shift in the Nakanishi-Lautrup auxiliary fields 
 \begin{equation}
 b_1 \rightarrow b_2-\frac{1}{2} [\bar c_2, c_2], \ \  \tilde b_1 \rightarrow \tilde b_2-\frac{1}{2} [\tilde{\bar c_2}, \tilde c_2].
 \end{equation}
The BRST transformation under which the effective action in non-linear gauge
remains invariant is given by  
\begin{eqnarray}
\delta_b \,\Gamma_{a} = \nabla_a    c_2\ \Lambda, && \delta_b\, \tilde\Gamma_{a} =\tilde\nabla_a    
 \tilde c_2\ \tilde\Lambda, \nonumber \\
\delta_b \,c_2 = -\frac{1}{2} {[c_2,c_2]}_ { }\ \Lambda, && \delta_b \,\tilde{ {c}}_2 = - \frac{1}{2}  [\tilde{ {c}}_2 ,  \tilde c_2]_{ } \ \tilde\Lambda, \nonumber \\
\delta_b \,\bar{c}_2 = b_2\ \Lambda -\frac{1}{2}[\bar c_2, c_2] \Lambda, && \delta_b \,\tilde {\bar c}_2 = \tilde b_2\ \tilde\Lambda -\frac{1}{2}[\tilde {\bar c}_2, \tilde c_2]\ \tilde\Lambda, \nonumber \\ 
\delta_b \,b_2 =-\frac{1}{2}[c_2, b_2]\Lambda- \frac{1}{8}[  [c_2,c_2] , \bar c_2]\Lambda, &&\delta_b \, \tilde b_2= -\frac{1}{2}[\tilde c_2, \tilde b_2] \tilde\Lambda- \frac{1}{8}[ [\tilde c_2, \tilde c_2], \tilde{\bar c_2}]\tilde\Lambda,  \nonumber \\ 
\delta_b \, X^{I } = i c_2   X^{I } \Lambda -  iX^{I }  \tilde c_2\ \tilde\Lambda, 
 &&  \delta_b \, X^{I \dagger }
 = i   \tilde c_2    X^{I \dagger }\ \tilde\Lambda - i  X^{I \dagger }  c_2\ \Lambda, 
\nonumber \\
\delta_b \, Y^{I } = i  \tilde c_2   Y^{I }\ \tilde\Lambda -iY^{I }   c_2\ \Lambda,  &&  
\delta_b \, Y^{I \dagger } = i c_2   Y^{I \dagger } \ \Lambda-  i Y^{I \dagger }
  \tilde c_2\ \tilde\Lambda.
  \label{brstnl}
\end{eqnarray}
 The effective action is also found invariant under the
 another set of BRST symmetry where roles of ghost and anti-ghost fields are interchanged,
 called as anti-BRST transformation and given by
\begin{eqnarray}
\delta_{ab} \,\Gamma_{a} = \nabla_a    \bar c_2\ \bar\Lambda, && \delta_{ab}\, \tilde\Gamma_{a} =\tilde\nabla_a    
 \tilde{\bar c_2}\ \tilde{\bar \Lambda}, \nonumber \\
\delta_{ab} \,\bar c_2 = - \frac{1}{2}{[\bar c_2, \bar c_2]}_ { }\ \bar\Lambda, &&\delta_{ab} \,\tilde{\bar {c}}_2 = -  \frac{1}{2}  [\tilde{\bar {c}}_2 ,  \tilde {\bar c}_2]_{ }\ \tilde{\bar \Lambda}, \nonumber \\
\delta_{ab} \, {c}_2 = -b_2\ \bar\Lambda-\frac{1}{2}[\bar c_2, c_2]\ \bar\Lambda, &&\delta_{ab} \,\tilde {  c}_2 = -\tilde b_2\ \tilde{\bar \Lambda} -\frac{1}{2}[\tilde {\bar c}_2, \tilde c_2]\ \tilde{\bar \Lambda}, \nonumber \\ 
\delta_{ab} \,b_2 =-\frac{1}{2}[  \bar c_2, b_2]\ \bar\Lambda +\frac{1}{8}[   [\bar c_2, \bar c_2], c_2]\ \bar\Lambda, &&\delta_{ab} \, \tilde b_2= -\frac{1}{2}
[ \tilde {\bar c}_2, \tilde b_2]\ \tilde{\bar \Lambda}+\frac{1}{8}  [[\tilde {\bar c}_2, \tilde {\bar c}_2], \tilde{  c}_2]\ \tilde{\bar \Lambda},  \nonumber \\ 
\delta_{ab} \, X^{I } = i \bar c_2   X^{I }\ \bar\Lambda -  iX^{I }  \tilde {\bar c}_2\ \tilde{\bar \Lambda}, 
 && \delta_{ab} \, X^{I \dagger }
 = i   \tilde {\bar c}_2    X^{I \dagger }\ \tilde{\bar \Lambda} - i  X^{I \dagger } \bar c_2\ \bar\Lambda, 
\nonumber \\
\delta_{ab} \, Y^{I } = i  \tilde{\bar c}_2   Y^{I } \ \tilde{\bar \Lambda}-iY^{I } \bar  c_2\ \bar\Lambda,  &&  
\delta_{ab} \, Y^{I \dagger } = i \bar c_2   Y^{I \dagger }\ \bar\Lambda -  i Y^{I \dagger }
  \tilde{\bar c}_2\ \tilde{\bar \Lambda}.
\end{eqnarray}
The above BRST and anti-BRST transformations are nilpotent as well as absolutely anticommuting, i.e. 
\begin{eqnarray}
\delta_{b}^2=0,\ \ \delta_{ab}^2 =0,\ \ \delta_{b}\delta_{ab} +\delta_{ab} \delta_{b}=0.
\end{eqnarray}
The gauge-fixing and ghost terms of the ABJM model in non-linear gauge
can be expressed in terms of BRST and anti-BRST exact terms as follows
\begin{eqnarray}
{\cal L}_{NL} &=&\frac{i}{2}\delta_{b}\delta_{ab}\int d^2\theta\   \mbox{Tr} \left[\Gamma_a \Gamma^a -\tilde \Gamma_a\tilde\Gamma^a -i\alpha
\bar c_2 c_2 +i\alpha \tilde{\bar c_2}\tilde c_2\right]_|,\nonumber\\
&=&-\frac{i}{2} \delta_{ab} \delta_{b}\int d^2\theta\  \mbox{Tr}\left[\Gamma_a \Gamma^a -\tilde \Gamma_a\tilde\Gamma^a -i\alpha
\bar c_2 c_2 +i\alpha \tilde{\bar c_2}\tilde c_2\right]_|.
\end{eqnarray}
In the next section we analyze the theory in BV formulation. 
\section{ ABJM theory in BV formulation}
To establish the theory in BV formulation we need to introduce antifields corresponding to fields with opposite statistics. In terms of fields/antifields, the generating functional for the ABJM theory in Lorentz-type gauge is,
\begin{eqnarray}
Z_L  &=&\int {\cal D} \Phi\ e^{iW_{L}[ \Phi , \Phi^\star, \tilde\Phi, \tilde\Phi^\star]}=\int {\cal D} \Phi \exp\left[i \int dv\,  \left({\cal L}_c + \int d^2\theta\  \mbox{Tr}\left[ \Gamma^{a\star} \nabla_a c_1 \right.\right.\right.\nonumber\\
&+&\left.\left.\left. \tilde{\Gamma}^{a\star} \tilde{\nabla}_a\tilde{c}_1 + \bar{c}_1^\star b_1 + \tilde{\bar{c}}_1^\star \tilde b_1
\right]\right)_|\right],
\label{zlin}
\end{eqnarray}
where $W_L$ is the extended quantum action and integration $\int dv$ refers to $\int d^3x$. 
The gauge-fixed fermion for ABJM theory in Lorentz gauge is defined by,
\begin{eqnarray}
\Psi_L = \bar{c}_1\left( iD^a \Gamma_a + \frac{\alpha}{2} b_1\right) - \tilde{\bar{c}}_1\left( iD^a\tilde{\Gamma}_a + \frac{\alpha}{2} \tilde{b}_1 \right). 
\end{eqnarray}
With the help of this gauge-fixed fermion we compute the antifields  for the Lorentz gauge as following: 
\begin{eqnarray} 
X^{I\star} &=& \frac{\delta \Psi_L}{\delta X^I} = 0, \, \, \, \, \, \, 
X^{I\dagger\star} = \frac{\delta \Psi_L}{\delta X^{I\dagger}} = 0, \, \, \, \, \, \,
Y^{I\star} = \frac{\delta \Psi_L}{\delta Y^{I}} = 0, \nonumber \\
Y^{I\dagger\star} &=& \frac{\delta \Psi_L}{\delta Y^{I\dagger}} = 0, \nonumber \, \, \, \, \, \, c_1^\star =  \frac{\delta \Psi_L}{\delta c_1} = 0, \, \, \, \, \, \,
\tilde{c}_1^\star =  \frac{\delta \Psi_L}{\delta \tilde{c}_1} = 0, \nonumber \\
\Gamma^{a\star}  &=& \frac{\delta \Psi_L}{\delta \Gamma_a} = - i D^a\bar{c}_1, \, \, \, \, \, \, \, \, \, \, \, \, \, \, \, \, \, \, \, \, \, \, \,
\tilde{\Gamma}^{a\star}  = \frac{\delta \Psi_L}{\delta \tilde{\Gamma}_a} = i D^a\tilde{\bar{c}}_1,\nonumber \\ 
\bar{c}_1^\star &=&  \frac{\delta \Psi_L}{\delta \bar{c}_1} =  iD^a \Gamma_a + \frac{\alpha}{2} b_1, \, \, \, \, \, \, \, \, 
\tilde{\bar{c}}_1^\star =  \frac{\delta \Psi_L}{\delta \tilde{ \bar{c}}_1} =  -iD^a \tilde{\Gamma}_a - \frac{\alpha}{2} \tilde{b}_1. 
\label{antil}
\end{eqnarray}
However, the generating functional for ABJM in the non-linear gauge in terms of fields/antifields is given by,
\begin{eqnarray}
Z_{NL}  &=&\int {\cal D} \Phi e^{iW_{NL}[ \Phi , \Phi^\star, \tilde\Phi, \tilde\Phi^\star]} =\int {\cal D} \Phi\ \exp   \left[i\int dv\, \left( {\cal L}_c + \int d^2\theta\ \mbox{Tr}\left[\Gamma^{a\star} \nabla_a c_2 + \tilde{\Gamma}^{a\star} \tilde{\nabla}_a\tilde{c}_2 \right.\right.\right. \nonumber\\
&+&\left.\left.\left.  \bar{c}_2^\star \left(b_2-\frac{1}{2}[\bar{c}_2,c_2]\right)+\tilde{\bar{c}}_2^\star \left (\tilde{b}_2-\frac{1}{2}[\tilde{\bar{c}}_2,\tilde{c}_2]\right)\right]\right)_| 
\right].
\label{znil}
\end{eqnarray}
We evaluate the expression for the gauge-fixing fermion for the non-linear gauge as following:
\begin{eqnarray} 
\Psi_{NL}  = \bar{c}_2 \left( iD^a\Gamma_a + \frac{\alpha}{2}b_2 - \frac{\alpha}{4} [\bar{c}_2,c_2] \right) -\tilde {\bar{c}}_2 \left( iD^a\tilde{\Gamma}_a + \frac{\alpha}{2}\tilde{b}_2- \frac{\alpha}{4} [\tilde{\bar{c}}_2,\tilde{c}_2] \right).
\end{eqnarray}
The antifields in this case are identified as,
\begin{eqnarray} 
X^{I\star} &=& \frac{\delta \Psi_{NL}}{\delta X^I} = 0, \, \, \, \, \, \, 
X^{I\dagger\star} = \frac{\delta \Psi_{NL}}{\delta X^{I\dagger}} = 0, \, \, \, \, \, \,
Y^{I\star} = \frac{\delta \Psi_{NL}}{\delta Y^{I}} = 0, \nonumber \\
Y^{I\dagger\star} &=& \frac{\delta \Psi_{NL}}{\delta Y^{I\dagger}} = 0, \nonumber \, \, \, \, \, \, c_2^\star =  \frac{\delta \Psi_{NL}}{\delta c_2} = 0, \, \, \, \, \, \,
\tilde{c}_2^\star =  \frac{\delta \Psi_{NL}}{\delta \tilde{c}_2} = 0, \nonumber \\
\Gamma^{a\star}  &=& \frac{\delta \Psi_{NL}}{\delta \Gamma_a} = - i D^a\bar{c}_2, \, \, \, \, \, \, \, \, \, \, \, \, \, \, \, \, \, \, \, \, \, \, \,
\tilde{\Gamma}^{a\star}  = \frac{\delta \Psi_{NL}}{\delta \tilde{\Gamma}_a} = i D^a\tilde{\bar{c}}_2,\nonumber \\ 
\bar{c}_2^\star &=&  \frac{\delta \Psi_{NL}}{\delta \bar{c}_2} = iD^a\Gamma_a + \frac{\alpha}{2}b_2 - \frac{\alpha}{4} [\bar{c}_2,c_2] , \, \, \, \, \, \, \, \, 
\tilde{\bar{c}}_2^\star =  \frac{\delta \Psi_{NL}}{\delta \tilde{ \bar{c}}_2} =  -iD^a\tilde{\Gamma}_a - \frac{\alpha}{2}\tilde{b}_2+ \frac{\alpha}{4} [\tilde{\bar{c}}_2,\tilde{c}_2].
\label{antinl}
\end{eqnarray}
We note the difference between the two extended quantum actions as follows,
\begin{eqnarray} 
 W_{NL} -  W_{L}  &=&   \int dv\int d^2\theta \ \mbox{Tr} \bigg[ \bigg( -iD^a \bar{c}_2\nabla_ac_2  + iD^a \bar{c}_1\nabla_a c_1 + i D^a \Gamma_a (b_2 -b_1) - \frac{i}{2} D^a \Gamma_a [\bar{c}_2,c_2]  \nonumber \\  &+& \frac{\alpha}{2} (b_2^2 - b_1^2) +  \frac{\alpha}{8} [\bar{c}_2,c_2]^2 - \frac{\alpha}{2}b_2 [\bar{c}_2,c_2] \bigg) + \bigg(iD^a \tilde{\bar{c}}_2\nabla_a\tilde{c}_2  - iD^a \tilde{\bar{c}}_1\nabla_a \tilde{c}_1 \nonumber \\ &-& i D^a \tilde{\Gamma}_a (\tilde{b}_2 -\tilde{b}_1)
+  \frac{i}{2} D^a \tilde{\Gamma}_a [\tilde{\bar{c}}_2,\tilde{c}_2]- \frac{\alpha}{2} (\tilde{b}_2^2 - \tilde{b}_1^2) - \frac{\alpha}{8} [\tilde{\bar{c}}_2,\tilde{c}_2]^2 +\frac{\alpha}{2} \tilde{b}_2 [\tilde{\bar{c}}_2, \tilde{c}_2]\bigg) \bigg]_|.
\label{diff}
\end{eqnarray}
The extended quantum actions, $W_{\Psi}[\Phi,\Phi^\star]\equiv (W_{NL}, W_{L})$, satisfies  certain rich mathematical
relation so-called  quantum master equation,  which is given by
\begin{equation}
\Delta e^{iW_{\Psi }[\Phi,\Phi^\star] } =0,\ \
 \Delta\equiv \frac{\partial_r}{
\partial\Phi^\star}\frac{\partial_l}{\partial\Phi } (-1)^{\epsilon
+1}.
\label{mq}
\end{equation}
Here we note that the extended quantum actions $W_{NL}$ and $W_{L}$ are two
different possible solutions of the quantum master equation.

In the next section, our goal would be to establish a map between the two generating functionals
corresponding to the above extended actions using the technique of field/antifield dependent BRST transformations.
\section{A mapping between solutions of quantum master equation}
We first analyze the  field/antifield dependent BRST transformation which is characterized by the field/antifield dependent BRST  parameter.
To achieve the goal, we   define the usual BRST transformation for the generic fields $\Phi_\alpha(x)$ and $\tilde\Phi_\alpha(x)$ written compactly as
 \begin{eqnarray}
\Phi_\alpha'(x)-\Phi_\alpha(x)&=&\delta_b  \Phi_\alpha(x)= s_b  \Phi_\alpha(x)\Lambda ={\cal R}_\alpha(x) \Lambda,\nonumber\\
\tilde\Phi_\alpha'(x)-\tilde\Phi_\alpha(x)&=&\delta_b \tilde \Phi_\alpha(x)= s_b  \tilde\Phi_\alpha(x)\tilde\Lambda =\tilde{\cal R}_\alpha(x) \tilde\Lambda,
 \end{eqnarray}
where ${\cal R}_\alpha(x)(s_b  \Phi_\alpha(x))$ and $\tilde{\cal R}_\alpha(x)(s_b  \tilde\Phi_\alpha(x))$ are the  Slavnov variations of the field $\Phi_\alpha(x)$ and $\tilde\Phi_\alpha(x)$ 
satisfying $\delta_b {\cal R}_\alpha(x)=\delta_b \tilde{\cal R}_\alpha(x) =0$.
Here the infinitesimal transformation  parameters $\Lambda$  and $\tilde\Lambda$ are  the  Grassmann parameters
and don't depend on any field/antifield.

Now, we present the  field/ antifield dependent BRST transformation   as follows
 \begin{eqnarray}
\delta_b  \Phi_\alpha(x)&=&\Phi_\alpha'(x)-\Phi_\alpha(x)={\cal R}_\alpha(x) \Lambda [\Phi,\Phi^\star],\nonumber\\
\delta_b  \tilde\Phi_\alpha(x)&=&\tilde\Phi_\alpha'(x)-\tilde\Phi_\alpha(x)=\tilde{\cal R}_\alpha(x) \tilde\Lambda [\tilde\Phi,\tilde\Phi^\star],\label{qg}
 \end{eqnarray}
 where the Grassmann parameters   $\Lambda [\Phi,\Phi^\star]$ and $\tilde\Lambda [\tilde\Phi,\tilde\Phi^\star]$ depend  on the field/antifield explicitly. The field/antifield dependent BRST transformation for the ABJM theory is constructed by making the transformation parameter of (\ref{brstl}) and (\ref{brstnl}) field/antifield dependent. Though being symmetry of the extended action such field/antifield dependent transformation is not nilpotent any more. We notice that under such transformation the path integral measure of generating functional changes non-trivially.  We compute the
 the change in the generating functional  
 as follows,
 \begin{eqnarray} 
 \delta_b Z_{L}  &=& \int {\cal D} \Phi ( \text{sDet} J[ \Phi , \Phi^\star, \tilde\Phi, \tilde\Phi^\star] )e^{i W_{L}[ \Phi , \Phi^\star, \tilde\Phi, \tilde\Phi^\star]}, \nonumber \\
 &=& \int {\cal D} \Phi e^{i (W_{L}[ \Phi , \Phi^\star, \tilde\Phi, \tilde\Phi^\star]-i \text{sTr} \ln J[\Phi, \Phi^\star])}.
 \label{deltaZ}
 \end{eqnarray}
 Furthermore, the Jacobian matrix appearing above for the field/antifield dependent BRST transformation is given by
 \begin{eqnarray}
 J_\alpha^{\, \, \beta} [ \Phi , \Phi^\star, \tilde\Phi, \tilde\Phi^\star] = \frac{ (\delta \Phi'_\alpha, \delta \tilde\Phi'_\alpha)}{(\delta \Phi_\beta, \delta \tilde\Phi_\beta)} &=& \delta_\alpha^{\, \, \beta}  + \frac{\delta {\cal R}_\alpha(x) }{\delta \Phi_\beta} \Lambda [\Phi, \Phi^\star] + {\cal R}_\alpha(x) \frac{\delta \Lambda [ \Phi, \Phi^\star]}{\delta \Phi_\beta}\nonumber\\
 &+& \frac{\delta \tilde{\cal R}_\alpha(x) }{\delta \tilde\Phi_\beta} \tilde\Lambda [\tilde\Phi, \tilde\Phi^\star] + \tilde{\cal R}_\alpha(x) \frac{\delta \tilde\Lambda [\tilde \Phi, \tilde\Phi^\star]}{\delta \tilde\Phi_\beta}.
 \label{det} 
 \end{eqnarray} 
 Utilizing (\ref{det})  and the nilpotency  of the BRST transformation (i.e. $s_b^2 = 0$) we obtain the following relation \cite{lav}
 \begin{eqnarray} 
 \text{sTr} \ln J[ \Phi , \Phi^\star, \tilde\Phi, \tilde\Phi^\star] = -\ln ( 1+ s_b \Lambda [\Phi, \Phi^\star ]+ s_b \tilde\Lambda [\tilde\Phi, \tilde\Phi^\star ] ).
 \label{jaco}
\end{eqnarray} 
 Because of the anticommuting nature of $\Lambda[\Phi, \Phi^\star]$ the determinant  simplifies to 
\begin{eqnarray} 
\text{sDet} J[ \Phi , \Phi^\star, \tilde\Phi, \tilde\Phi^\star] = \frac{1}{1  + s_b \Lambda[\Phi, \Phi^\star]+ s_b \tilde\Lambda [\tilde\Phi, \tilde\Phi^\star ]}.
\end{eqnarray}
Plugging this value of   determinant in the relation (\ref{deltaZ}) we get
\begin{eqnarray} 
 s_b Z_{L}   = \int {\cal D} \Phi \exp   \bigg(iW_{L}[ \Phi , \Phi^\star, \tilde\Phi, \tilde\Phi^\star]- \ln ( 1+ s_b \Lambda [\Phi, \Phi^\star ]+ s_b \tilde\Lambda [\tilde\Phi, \tilde\Phi^\star ] )\bigg).
 \label{deltabZ}
 \end{eqnarray}
This is a very general expression for the change in the generating functional of the ABJM theory under field/antifield dependent BRST transformation because it involves an arbitrary $\Lambda[\Phi,\Phi^\star]$.
Now we evaluate such variation under an specific choice of the field/antifield dependent transformation parameters chosen as  follows
\begin{eqnarray}
\Lambda[\Phi, \Phi^\star] &=& \int dv\int d^2\theta\ \psi (s_b   \psi)^{-1} \bigg( \exp \bigg\{ -i s_b  \psi \bigg\} -1 \bigg)_|,\nonumber\\
\tilde\Lambda[\tilde\Phi, \tilde\Phi^\star] &=& \int dv\int d^2\theta\ \tilde\psi (s_b   \tilde\psi)^{-1} \bigg( \exp \bigg\{ -i s_b  \tilde\psi \bigg\} -1 \bigg)_|,
\label{res1}
\end{eqnarray}
where $\psi$ and $\tilde\psi$ are defined by
\begin{eqnarray}
 \psi &= & ( \bar{c}_2 \bar{c}_2^\star - \bar{c}_1 \bar{c}_1^\star ),\nonumber\\
 \tilde \psi &= &   ( \tilde{\bar{c}}_2 \tilde{\bar{c}}_2^\star -\tilde{\bar{c}}_1 \tilde{ \bar{c}}_1^\star ) .
 \label{res2}
 \end{eqnarray}
 We now demonstrate that the above choice of $\Lambda$ and $\tilde \Lambda$ relate the two generating functionals (\ref{zlin}) and (\ref{znil}). This is one of the main results of this paper. \\
 The Jacobian  expression (\ref{jaco}) for the above choice of parameter yields,
 \begin{eqnarray} 
 i \ln ( 1 + s_b \Lambda [\Phi, \Phi^\star ] + s_b \tilde\Lambda [\tilde\Phi, \tilde\Phi^\star ]) &=&\int dv\int d^2\theta\ (s_b \psi +s_b\tilde{\psi}) \nonumber \\ 
 &=&\int dv\int d^2\theta\left[  (s_b \bar{c}_2)  \bar{c}_2^\star - (s_b \bar{c}_1) \bar{c}_1^\star  +\bar{c}_2 (s_b \bar{c}_2^\star) - \bar{c}_1 (s_b\bar{c_1}^\star)   \right. \nonumber \\ 
 &+  &\left.   (s_b \tilde{\bar{c}}_2) \tilde{ \bar{c}}_2^\star - (s_b \tilde{\bar{c}}_1) \tilde{\bar{c}}_1^\star +\tilde{\bar{c}}_2 (s_b \tilde{\bar{c}}_2^\star) - \tilde{\bar{c}}_1 (s_b\tilde{\bar{c}}_1^\star)\right]_|. 
 \label{sb}
 \end{eqnarray} 
Now we can use the antifield expressions (\ref{antil}), (\ref{antinl}) and the linear and non-linear BRST transformations (\ref{brstl}), (\ref{brstnl}) to complete the computation. There are eight terms in the   parentheses, let us calculate some of them. Firstly, we calculate
\begin{eqnarray}
(s_b \bar{c}_2)  \bar{c}_2^\star  &=& \bigg( b_2 -\frac{1}{2} [\bar{c}_2,c_2]\bigg) \bigg( iD^a \Gamma_a +\frac{\alpha}{2}b_2- \frac{\alpha}{4} [\bar{c}_2,c_2]\bigg), \nonumber \\ 
&=&  b_2\bigg(iD^a\Gamma_a +\frac{\alpha}{2}b_2\bigg) - \frac{i}{2}D^a\Gamma_a [\bar{c}_2,c_2] +\frac{\alpha}{8} [\bar{c}_2,c_2]^2 - \frac{\alpha}{2}b_2 [\bar{c}_2,c_2] .
\label{35}
\end{eqnarray}
The second term leads to 
\begin{eqnarray}
 (s_b \bar{c}_1) \bar{c}_1^\star &=& b_1\bigg(iD^a\Gamma_a +\frac{\alpha}{2}b_1\bigg).
 \label{36}
\end{eqnarray}
However, the third term is computed as,
 \begin{eqnarray}
 \bar{c}_2 (s_b \bar{c}_2^\star) &=& \bar{c}_2 \bigg( iD^a s_b\Gamma_a +\frac{\alpha}{2}s_b b_2+ \frac{\alpha}{4} s_b[\bar{c}_2,c_2]\bigg).
 \label{37}
 \end{eqnarray}
 Now, utilizing the Slavnov variation of (\ref{brstnl}) we have,
 \begin{eqnarray} 
 && s_b \bar{c}_2  = b_2 -\frac{1}{2} [\bar{c}_2,c_2], \nonumber \\
 && \mbox{or,  } s_b^2 \bar{c}_2 =  0 = s_b b_2 -\frac{1}{2} s_b   [\bar{c}_2,c_2], \nonumber \\
&&\mbox{or,  } s_b   [\bar{c}_2,c_2] = 2 s_b b_2.\label{vh}
 \end{eqnarray}
 Putting the values of (\ref{vh}) back in (\ref{37}) gives
  \begin{eqnarray}
 \bar{c}_2 (s_b \bar{c}_2^\star) &=& \bar{c}_2 \bigg( iD^a s_b\Gamma_a +\frac{\alpha}{2}s_b b_2- \frac{\alpha}{4} s_b[\bar{c}_2,c_2]\bigg), \nonumber \\ 
 &=&  \bar{c}_2 \bigg( iD^a s_b\Gamma_a +\frac{\alpha}{2}s_b b_2- \frac{\alpha}{4} (2 s_b b_2)\bigg), \nonumber \\
 &=& \bar{c}_2 i D^a \nabla_a c_2 = -i D^a \bar{c}_2 \nabla_a c_2.
 \label{39}
 \end{eqnarray}
The fourth term is calculated by,
 \begin{eqnarray} 
 \bar{c}_1 s_b \bar{c}_1^\star = \bar{c}_1 i D^a \nabla_a c_1  = -i D^a \bar{c}_1 \nabla_a c_1
 \label{40}
 \end{eqnarray}
 Putting together (\ref{35}), (\ref{36}), (\ref{39}) and (\ref{40}) we obtain
 the following expression
 \begin{eqnarray}
  (s_b \bar{c}_2)  \bar{c}_2^\star - (s_b \bar{c}_1) \bar{c}_1^\star  +\bar{c}_2 (s_b \bar{c}_2^\star) - \bar{c}_1 (s_b\bar{c_1}^\star) &=&  -iD^a \bar{c}_2\nabla_ac_2  + iD^a \bar{c}_1\nabla_a c_1 + i D^a \Gamma_a (b_2 -b_1)  \nonumber \\
   &-&   \frac{i}{2} D^a \Gamma_a [\bar{c}_2,c_2]+\frac{\alpha}{2} (b_2^2 - b_1^2) 
+\frac{\alpha}{8} [\bar{c}_2,c_2]^2 \nonumber\\
& - &\frac{\alpha}{2}b_2 [\bar{c}_2,c_2].
\label{1}
\end{eqnarray}
Following a similar computation we have for
\begin{eqnarray} 
 (s_b \tilde{\bar{c}}_2) \tilde{ \bar{c}}_2^\star - (s_b \tilde{\bar{c}}_1) \tilde{\bar{c}}_1^\star + \tilde{\bar{c}}_2 (s_b \tilde{\bar{c}}_2^\star) - \tilde{\bar{c}}_1 (s_b\tilde{\bar{c}}_1^\star)  &=&   iD^a \tilde{\bar{c}}_2\nabla_a\tilde{c}_2  - iD^a \tilde{\bar{c}}_1\nabla_a \tilde{c}_1 - i D^a \tilde{\Gamma}_a (\tilde{b}_2 -\tilde{b}_1)
 \nonumber \\ 
 &+& \frac{i}{2} D^a \tilde{\Gamma}_a [\tilde{\bar{c}}_2,\tilde{c}_2]-\frac{\alpha}{2} (\tilde{b}_2^2 - \tilde{b}_1^2)  - \frac{\alpha}{8} [\tilde{\bar{c}}_2,\tilde{c}_2]^2\nonumber\\
 & +& \frac{\alpha}{2} \tilde{b}_2 [\tilde{\bar{c}}_2, \tilde{c}_2].
 \label{2}
\end{eqnarray}
Therefore, it is easy to see from the equations (\ref{diff}),  (\ref{deltabZ}),(\ref{sb}),(\ref{1}) and (\ref{2}) that 
\begin{eqnarray} 
 \delta_b Z_L = Z_{NL}.
 \end{eqnarray}
 Hence we have shown that under field/antifield dependent BRST transformation with the appropriate
 choice of parameters (\ref{res1}) and (\ref{res2}), the different solutions of the quantum master equation can be related.
\section{conclusion}
In this paper we have 
established the ABJM theory at quantum level by investigating it in the BV formulation on $\mathcal{N}=1$ superspace. For this purpose, we have extended the configuration space by
introducing the antifields corresponding to the fields of ABJM model.
Further, we have calculated the exact values of antifields by
choosing the suitable gauge-fixing fermion. 
We have mainly discussed the Lorentz-type and Curci-Ferrari type gauges 
from the BRST quantization perspectives. The quantum master equation for the ABJM theory, having different possible solutions, is also established. Furthermore, we have generalized the 
BRST symmetry of the theory by developing the field/antifield dependent parameters.
Here we need two parameters of transformation rather than one.
We have also successfully demonstrated how a particular choice of the transformation parameters can relate two different generating functionals in the Lorentz-type and the Curci-Ferrari type gauges.

Our analysis on BV formulation of ABJM theory will provide a convenient way to study
the possible violations of the symmetries of the action by quantum effects.
Such analysis may also be useful in calculating the $S$-matrix of the theory because  we have
already computed the definite values of antifields.  The master equation
discussed above is more fundamental than the Zinn-Justin equation which guarantees
the renormalizability  of the ABJM theory, since the master equation relies on the
fundamental action rather than the quantum effective action.
The present investigation is a step towards the study of the 
deformations of the action and anomalies.

\section*{Acknowledgments}
SU acknowledges the support from the CNPq-Brazil under the Grant No. 504542/2013-3. DD acknowledges the hospitality of the Kavli Institute for Theoretical Physics and NSF Grant PHY11-25915
for support. DD was supported by a KITP Graduate Fellowship.

\end{document}